# Determination of Thickness-dependent Damping Constant and Plasma Frequency for Ultrathin Ag and Au Films: Nanoscale Dielectric Function

Luis J. Mendoza-Herrera[a,b,1*], Myrian C. Tebaldi[a,b], Lucía B. Scaffardi[a] and Daniel C. Schinca[a,b]

[a] Centro de Investigaciones Ópticas (CIOp),(CONICET-CIC-UNLP), Camino Centenario y 506, Gonnet, 1897, La Plata, Argentina
[b] Facultad de Ingeniería, UNLP, 1 y 47, 1900, La Plata, Argentina
[1*] Corresponding author: joaquinm@ciop.unlp.edu.ar

**Abstract**

There is an ever increasing interest in the development of plasmonic 2D nanomaterials, with widespread applications in optoelectronics, high resolution microscopy, imaging and sensing, among others. With the current ability of ultrathin noble metal film deposition down to a few monolayers in thickness, there is a need for an analytical expression of the thickness dependent complex dielectric function for predicting optical properties for arbitrary thicknesses.

The free and bound electron contributions to the dielectric function are dealt with independently, since their influences affect separate wavelengths ranges. The former is dealt within the Drude model framework for large wavelengths with appropriately addressed damping constant and plasma frequency parameters to account for thickness dependence. Applying our previously developed method, we determine these parameters for specific film thicknesses, based on refractive index experimental values for Ag and Au thin films. Fitting separately each one of these parameters allowed us to find an analytical expression for their dependence on arbitrary film thickness and consequently for the free electron contribution.

Concerning bound electrons, it is seen that its contribution for small wavelengths is the same for all analyzed thicknesses and may be set equal to the bulk bound contribution.

Taking all these facts into account, the complex dielectric function can be rewritten analytically, in terms of the bulk dielectric function plus corrective film thickness dependent terms.

In particular, the fitting process for the damping constant allows us to determine that the electron scattering at the film boundary is mainly diffusive (inelastic) for both silver and gold thin films. It is also shown that, in accordance with theoretical studies, plasma frequency shows a red shift as the film thickness decreases.

1. **Introduction**

Nanoscale metal materials, such as nanoparticles (NPs), nanowires and thin films, have an ever increasing interest in optics and photonics, mainly due to their special optical properties in field enhancement, non linear optics and nanoantennas among others. Metal NPs were first studied for their applications in biology and medicine [1, 2]. Metal nanowires found applications in solar cells [3], while metal films are used in development of plasmonic sensors and plasmonic waveguides [4-10]. These applications are based on the distinctive optical properties at the nanoscale, which in turn, are based on knowledge of the size dependent dielectric function properties. Within the Drude model approach, it is important then to determine the size dependent damping constant and plasma frequency optical constants for studies related to photonics [11], plasmonics [12, 13], clusters of NPs [14] and characterization of nanostructures by Optical Extinction Spectroscopy [15-21]. Knowledge derived from these studies may be used for on-demand design of devices for specific applications.

Nowadays, the increasing ability in the precise control of film deposition and material composition allows the production of thin and ultrathin films, enhancing the application of 2D materials in different fields of science and technology. At the nanometric scale, the dependence of dielectric function on the dimensions of the material is mainly evidenced through the dependence of damping constant on mean free path. With respect to this subject, there are theoretical studies devoted to determining an adequate description of the mean free path in terms of size for spherical, cylindrical and cartesian (film) geometries [22-24]. Concerning the plasma frequency, it has traditionally been considered constant and independent of size of the nanoscale. However, for the case of ultrathin plasmonic films, Bondarev et al. [25] and Shah et al. [26] showed that plasma frequency is indeed dependent on thickness.

Damping constant and plasma frequency values for thin films may be obtained from experimental refractive index data derived from conventional spectroscopic ellipsometric measurements [27, 28]. There are many papers devoted to retrieving wavelength dependent refractive index data based on experimental spectroscopic ellipsometry of noble metals films of various thicknesses. Traditionally, Ag and Au film refractive index data in the range from 18 to 50 nm thicknesses, reported by Johnson and Christy [29], are frequently used for optical properties characterization. Zhao et.al. [30] determined refractive index values for Ag films with thicknesses in the range from 5 nm to 15 nm. Mc Peack el al. [30] studied 300 nm thickness Ag films as well as Au film refractive index for the same thicknesses [31]. More recently, Yakubovsky et al. [32,33] measured Au films refractive index in the range from 4 nm to 117 nm thicknesses. It is important to point out that in none of these papers, the dependence of the optical constants on film thickness is analyzed.

In the present work, the thickness dependence of the damping constant and plasma frequency for Ag and Au film is determined for the first time, based on experimental refractive index data. Using a semiclassical approach starting from the Drude model, we introduce appropriate expressions for thickness dependence of damping constant (through its dependence of electron mean free path) and plasma frequency, which allow predicting their value for arbitrary film thicknesses. Using these expressions, the thin film dielectric function is rewritten in terms of the bulk dielectric function plus films thickness dependent terms. In the literature there exists a gross approximation in the determination of the dielectric function using an additive constant to take into account bound electron contribution. We avoid this uncertainty by using an alternative approach based on the direct experimental measurement of $\varepsilon_{bulk}(\omega)$, which contains all the information of bounds electron contribution.

Besides, from the fitting of the thickness dependent damping constant using Fuchs expressions for the mean free path, it is possible to derive the proportion of the elastic electron scattering process at the film boundary.

2. Theoretical description. Dielectric Function Model

Metal bulk dielectric function can be described as the sum of two contributions: free and bound electrons:

$$\varepsilon_{bulk}(\omega) = \varepsilon_{bulk}^{free}(\omega) + \varepsilon_{bulk}^{bound}(\omega) \tag{1}$$

Free electron contributions can be written using the Drude model [34-37] as:

$$\varepsilon_{bulk}^{free}(\omega) = 1 - \frac{\omega_p^2}{\omega^2 + i\omega\gamma_{free}} \qquad (2)$$

where $\omega_p$ is the bulk material plasma frequency and $\gamma_{free}$ is the bulk damping constant. The former is defined as $\omega_p^2 = \frac{N/V}{m\varepsilon_0}e^2$ where $N$ is the amount of free electrons in a volume $V$, $m$ is the effective mass of an electron, and $e$ and $\varepsilon_0$ have the usual meaning. On the other hand, damping constant is written as $\gamma_{free} = v_F/l_\infty$, where $v_F$ is the Fermi velocity and $l_\infty$ is the electron mean free path for bulk material and accounts for the sum of electron-ion, electron-electron, electron-phonon and lattice defects scattering collisions.

Concerning bound electron contribution, it has been shown in previous works [11 – 13] that it can be determined, following Equation (1), as the difference between the experimental bulk dielectric function ($\varepsilon_{bulk}(\omega)$) and the free electron contribution calculated using Equation (2).

Confinement of a bulk material to nanometric scale in a determined direction, like in the case of thin films, causes an effective reduction in dimensionality, changing the way in which the confined electrons interact. From a semiclassical point of view, this fact comprises the modification of two parameters: the damping constant and the plasma frequency. Thickness dependence of the former can be readily taken into account by adding a size dependent term and can be expressed as:

$$\gamma_{free}^t(d) = \gamma_{free} + v_F/l_c \qquad (3)$$

where the superscript "$t$" indicates thickness dependent and the added term $(v_F / l_c)$ stands for the interaction between electrons and material boundary through the size-dependent electron mean free path $l_c$. This latter parameter will be discussed in section 3.1.1.

On the other hand, under this confinement condition, plasma frequency cannot be easily modified by adding a size dependent term to its bulk expression. For a metallic thin film of thickness $d$, Bondarev et al. [25] proposed an expression of the form :

$$\omega_p^t(d) = \omega_p f(d) \tag{4}$$

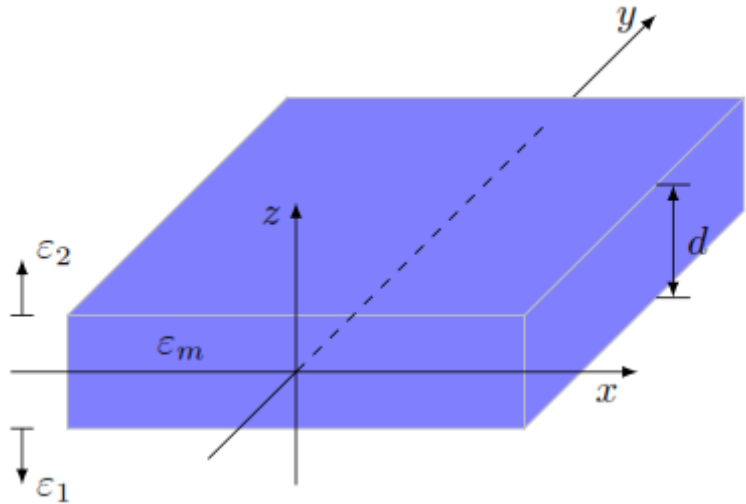

Figure 1: schematic of film thickness $d$ and dielectric constant $\varepsilon_m$, surrounded by media with dielectric constant $\varepsilon_1$ and $\varepsilon_2$

where the superscript "$t$" again indicates thickness dependent parameter. Based on the schematic of Figure 1, the authors show that for a metal film with permittivity $\varepsilon_m$ and z-dimension (thickness $d$) much smaller than those in-plane $x$ - $y$, deposited on a substrate of permittivity $\varepsilon_1$ and a superstrate of permittivity $\varepsilon_2$, with $\varepsilon_1$, $\varepsilon_2 \ll \varepsilon_m$ (or equivalently $\varepsilon_m$ $\gg \varepsilon_1 + \varepsilon_2$), the function $f(d)$ takes the form: $f(d) = 1/\sqrt{1 + (\varepsilon_1 + \varepsilon_2)/\varepsilon_m kd}$, where $k$ is the absolute value of the quasimomentum in bulk materials.

Considering Equation (3) and Equation (4), the thickness dependent free electron contribution may now be written as:

$$\varepsilon^t_{free}(\omega, d) = 1 - \frac{\left(\omega^t_p(d)\right)^2}{\omega^2 + i\omega\, \gamma^t_{free}(d)} \tag{5}$$

In the next section, we will determine $\gamma^t_{free}(d)$ and $\omega^t_p(d)$ from literature experimental values of thin film refractive index for different thicknesses, based on a method developed previously by our group and applied to spherical metal nanoparticles [18, 20]. Particularly, it will be shown that this approach allows determining $\gamma^t_{free}(d)$ (Equation 3), without the need of an explicit modeling of the mean free path of $l_c$.

## 2.1 Theoretical method for determination of $\gamma^t_{free}(d)$ and $\omega^t_p(d)$ from experimental thin film refractive index data

For sufficiently large wavelengths ($\lambda \gg$) (which depends on the considered metal), the dielectric function for a thin film can be described by the free electron (intraband transitions) Drude model [34-37], considering negligible bound electron contribution (interband transitions). Under this condition, Equation (5) reduces to:

$$\varepsilon^t(\omega, d)_{\lambda \gg} = \varepsilon^t_{free}(\omega, d) = 1 - \frac{\left(\omega^t_p(d)\right)^2}{\omega^2 + i\omega\, \gamma^t_{free}(d)} = \varepsilon'^{\,t}(\omega, d)_{\lambda \gg} + i\, \varepsilon''^{\,t}(\omega, d)_{\lambda \gg}$$

$$\tag{6}$$

where $\varepsilon'^{\,t}(\omega, d)_{\lambda \gg}$ and $\varepsilon''^{\,t}(\omega, d)_{\lambda \gg}$ represent the real and imaginary parts of the dielectric function. As it was shown in previous works [18,19], an adequate combination of these parts, yield two linear equations:

$$\omega\, \varepsilon''^{\,t}(\omega, d)_{\lambda \gg} = \gamma^{t}_{free}(d)\, F(\omega, d) \qquad (7)$$

$$G(\omega, d) = \left(\omega^{t}_{p}(d)\right)^{2} F(\omega, d) \qquad (8)$$

where

$$F(\omega, d) = 1 - \varepsilon'^{\,t}(\omega, d)_{\lambda \gg} \qquad (9)$$

and

$$G(\omega, d) = \omega^{2}\left[\left(\varepsilon''^{\,t}(\omega, d)_{\lambda \gg}\right)^{2} + \left(1 - \varepsilon'^{\,t}(\omega, d)_{\lambda \gg}\right)^{2}\right] \qquad (10)$$

Equations (7) and (8) allow to determine the values of damping constant $\gamma^{t}_{free}(d)$ and plasma frequency $\omega^{t}_{p}(d)$ respectively, from the slopes of the linear fits, when $\omega\, \varepsilon''^{\,t}(\omega, d)_{\lambda \gg}$ and $G(\omega, d)$ are plotted against $F(\omega, d)$. Determination of these parameters for the analyzed experimental film thicknesses allows us to build a fitting curve, with the aim to predict damping constant and plasma frequency for any thickness and determine the optical properties for any considered film.

## 3. Results

### 3.1. Thickness dependent damping constant for Ag films

The $\gamma_{free}^{t,Ag}(d)$ values for Ag films are obtained from Equations (7) to (10) by using the experimental refractive index values for the thicknesses reported by Zhao et. al. [30] ($d = $ 5.0, 6.20, 7.50, 9.80, 12.5 and 15.0 nm) as measured for a range of wavelengths from UV to NIR. Silver films were deposited by magnetron DC sputtering at room temperature. Film thicknesses were controlled stably by sputtering time in the sputtering process.

Figure 2 plots the relation described in Equation (7) for experimental data corresponding to 5.0 nm thickness. It can be observed that the data points show a linear region (pink dots) which can be fitted by an origin-crossing straight line (black full line) representing the Drude model given by Equation (7). Values of $F(\omega, d)$ for this linear region span from 28 to 220 approximately, that correspond to wavelengths between 0.84 µ$m$ and 2.5 µ$m$. The slope of this straight line yields a value of $\gamma_{free}^{t,Ag}(5.0nm) = 3.629 \times 10^{14} rad/s$. For wavelengths smaller than 0.84 µ$m$, the curve departs from linearity (green data points) due to the influence of interband transitions, not taken into account in the Drude model. However, for now, we will restrict our study to the linear region.

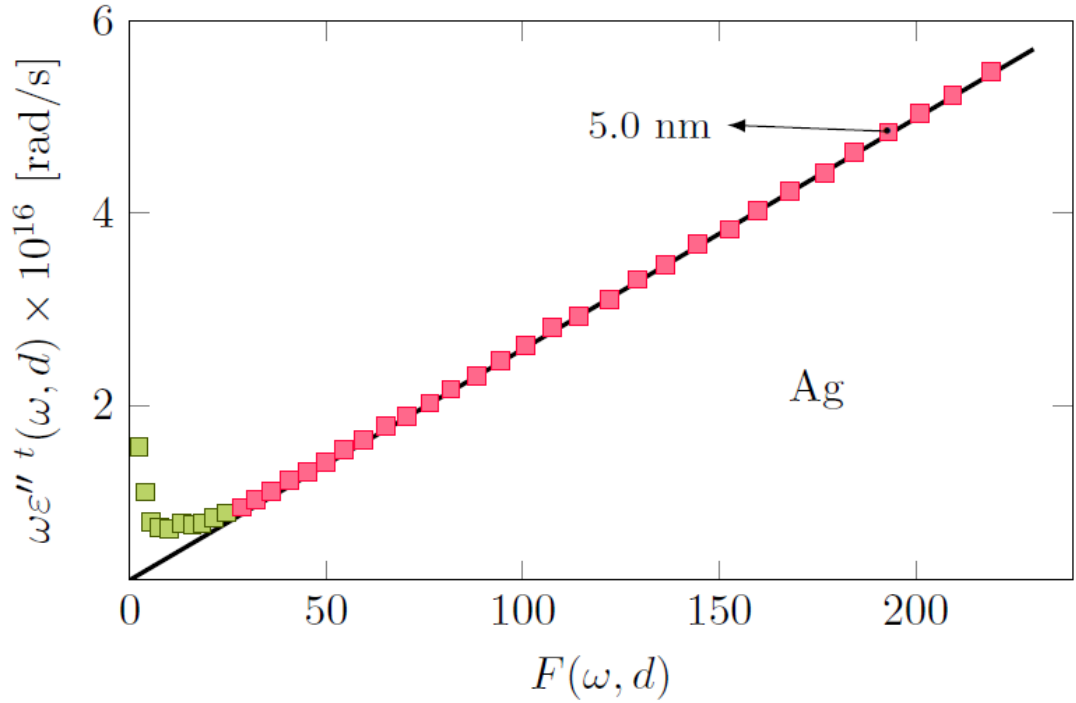

Figure 2: $\omega\varepsilon''^{t}(\omega, d)$ vs $F(\omega, d)$, for $d$ = 5.0 nm thickness Ag film. Pink dots were fitted with an origin-crossing linear fit (black full line) representing the Drude model given by Eq. (7).

This approach can be extended to the rest of Ag film thicknesses given by Zhao et al. [30]. Figure 3 shows these results, where again, the slope of the linear fit can be used to determine thickness-dependent damping constant. The data curve for 300 nm thickness film, calculated by using Mc Peak [31] refractive index data, is plotted in blue to show the linear fit for large thickness. The obtained values for $\gamma_{free}^{t,Ag}(d)$ are summarized in Table 1 together with the errors $\Delta\gamma_{free}^{t,Ag}(d)$ from the theoretical fits.

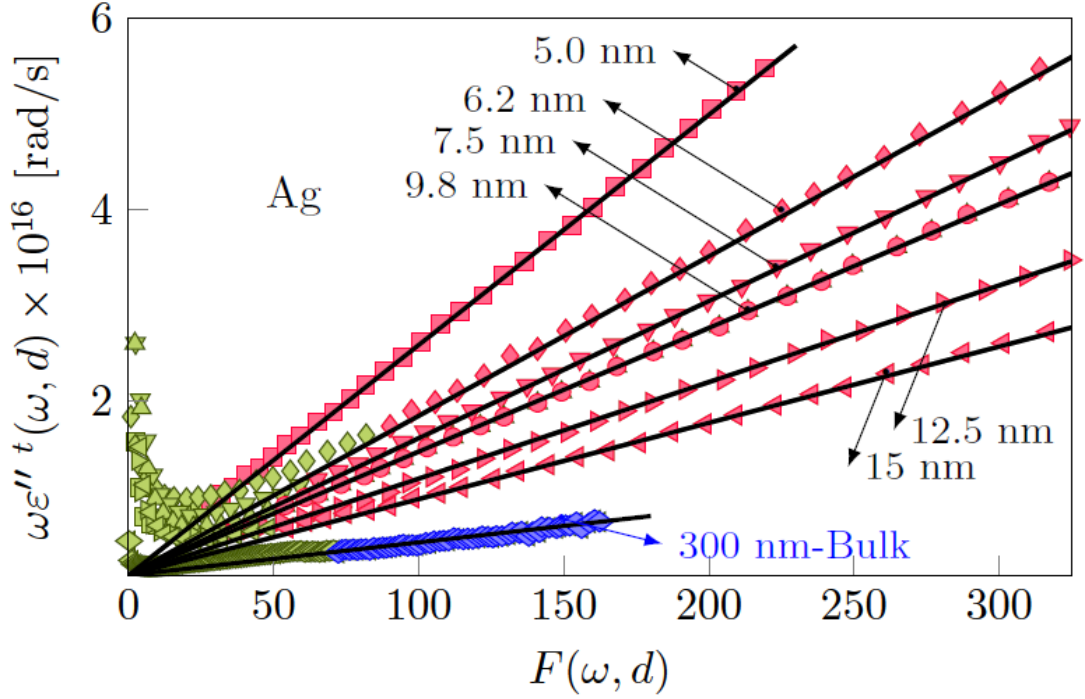

Figure 3. $\omega\,\varepsilon''(\omega, d)$ vs $F(\omega, d)$ for different Ag films thicknesses. From the slopes of the linear region, the damping constant can be determined. Pink symbols correspond to the linear region while green symbols lie outside the linear region. The numbers indicate the different thickness values ($d$).

Table 1: Calculated damping constant for different Ag film thicknesses.

| $d$ [nm] | $\gamma_{free}^{t,Ag}(d) \times 10^{14}$ [rad/s] | $\Delta\gamma_{free}^{t,Ag}(d) \times 10^{14}$ |
|---|---|---|
| 5.0 | 3.629 | ±0.518 |
| 6.2 | 1.897 | ±0.284 |
| 7.5 | 1.532 | ±0.216 |
| 9.8 | 1.379 | ±0.195 |
| 12.5 | 1.092 | ±0.154 |
| 15.0 | 0.887 | ±0.126 |
| 300 (Bulk) | 0.233 | ±0.084 |

Visual inspection of the first and second columns in Table 1, shows that with increasing film thickness, the damping constant decreases, seemingly approaching a limiting bulk value. The last line in Table 1 shows the damping constant value for 300 nm ($\gamma_{free}^{t,Ag}(300\ nm) = 0{,}233 \times 10^{14} rad/s$) which may be considered as a bulk limit. This fact will be discussed in the next section 3.1.1. This dependence of damping constant with film thickness is related to a reduction in electron mean free path $l_c$, since there is an increase in collision frequency with the film boundary in the z direction as the thickness decreases. In the next section we will discuss the mean free path together with the bulk value.

### 3.1.1. On the mean free path

For a graphical view of the relation between the damping constant and the film thickness shown in Table 1, Figure 4 plots the obtained values of $\gamma_{free}^{t,Ag}(d)$ vs $d$ (blue dots) together with the corresponding error bars. Recalling Equation (3) in section 2, and taking into account the additive term that depends inversely on $l_c$, it is possible to fit the data values in Figure 4 if a dependence of $l_c$ with $d$ is known. There are several theoretical classical approaches for determination of $l_c$ for different nanoparticle geometries [22-24, 34]. For thin films, Fuchs [24] derives an expression for the mean free path in the case of metal films considering partial elastic scattering:

$$l_c = l_\infty - \frac{3l_\infty(1-s)}{8\kappa} + \frac{3l_\infty(1-s)^2}{4\kappa} \sum_{j=1}^{\infty} s^{j-1}\left[B(\kappa j)\left(\kappa^2 j^2 - \frac{\kappa^4 j^4}{12}\right) + e^{-\kappa j}\left(\frac{1}{2} - \frac{5\kappa j}{6} - \frac{\kappa^2 j^2}{12} + \frac{\kappa^3 j^3}{12}\right)\right]$$

(11)

where $s$ is the proportion of elastically scattered electrons on the surface, $\kappa = d/l_\infty$, with $d$ being the film thickness and the function $B(\kappa j)$ defined as: $B(\kappa j) = \int_{\kappa j}^{\infty} \frac{e^{-x}}{x} dx$. The best fit (black full line) corresponds to $s = 0.25$ and $\gamma_{free}^{Ag} = 0.243 \times 10^{14} rad/s$, using $v_F = 1.39 \times 10^6 m/s$ [38]. Inset shows a plot enlargement for thicknesses $d < 20$ nm.

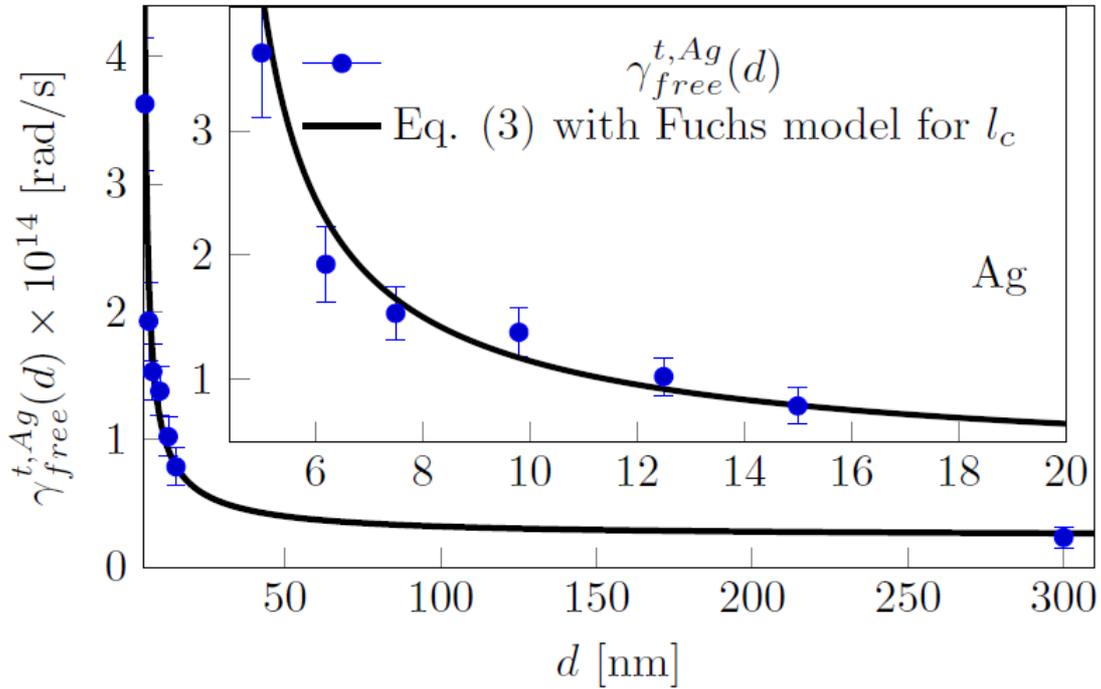

Figure 4. Damping constant in terms of the Ag film thickness. Blue dots are experimental data points. Black full line corresponds to a fit using Equation (3) together with Fuchs model for $l_c$

It can be seen that there is a very good agreement between the empirical data (blue dots) and the theoretical fit, suggesting that for these Ag film thicknesses, electron scattering at the film boundary is mainly diffusive (inelastic).

The fitting curve shows little variation for $d > 150$ nm, suggesting an asymptotic behavior whose limit is $\gamma_{free}^{Ag} = 0.243 \times 10^{14} rad/s$, when the Fuchs model is used. This value agrees, within calculated errors, with the value of the damping constant derived from the refractive index value obtained by Mc Peak (last line in Table 1). For this reason, we will consider from now on 300 nm as a bulk regime.

The electron mean free path for bulk Ag can be derived recalling the relation $\gamma_{free} = v_F / l_\infty$, obtaining a value of $l_\infty = 57$ nm, which is very close to 53 nm calculated by Gall [39] using ab initio formalism.

**3.2. Thickness dependent damping constant for Au films**

Similarly to the analysis carried out for Ag films, values of $\gamma_{free}^{t,Au}(d)$ were determined from Equations (7) to (10) by using the experimental refractive index values reported by Yakubovsky et al. [32, 33] for of controlled thin films thicknesses $d = 4.1, 6.1, 9.0, 25, 53$ and 117 nm. Films were deposited from high purity gold pellets on chemically pre-cleaned silicon Si (100) wafer by Electron Beam Evaporation (EBE) deposition technique. Figure 5 shows $\omega \varepsilon''(\omega, d)$ (Equation 7) vs $F(\omega, d)$ (Equation 9) for the different Au films thicknesses analyzed. The plots are presented in two distinct panels since the wavelength range for data collection is different for thicknesses smaller than 9 nm (a) and larger than 9 nm (b). In the latter panel, the data curve for 300 nm thickness film given by Mc Peak [31] is added in blue data dots to show the bulk limit.

Again, the slope of the linear fit can be used to determine damping constant. This linear region is represented by pink dots which can be described using Drude model. Shorter wavelength data (green symbols) depart from linearity showing the influence of interband transitions. From the slope of the fit of the linear region (black full line), damping constant can be determined for each film thickness. The obtained values of $\gamma_{free}^{t,Au}(d)$ are summarized in Table 2, together with the errors from the theoretical fits.

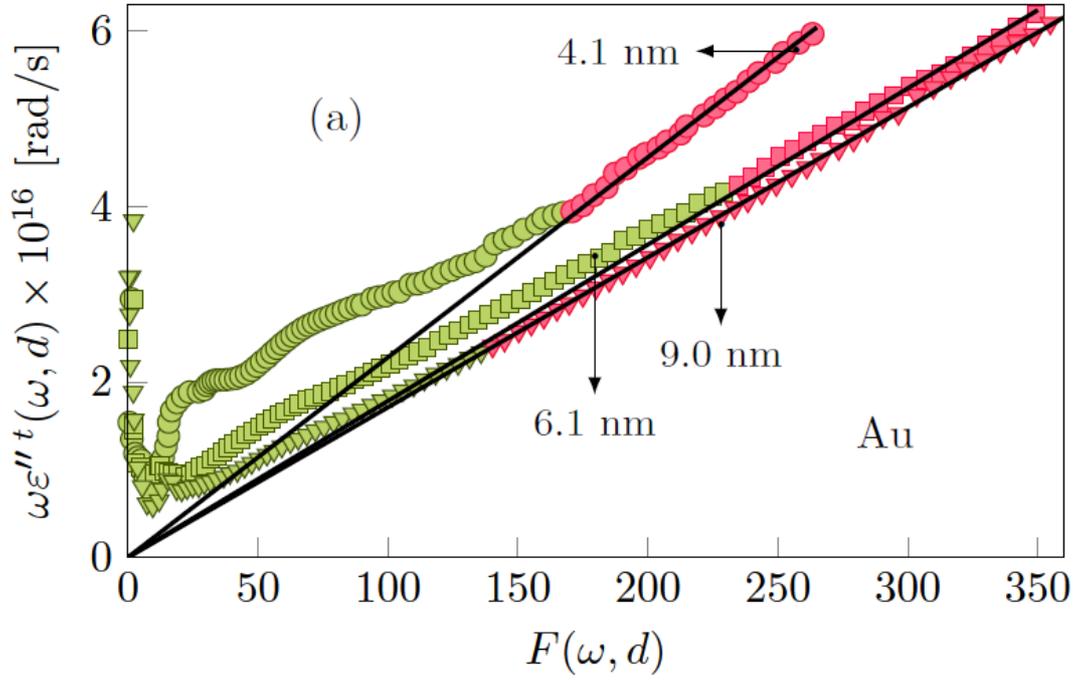

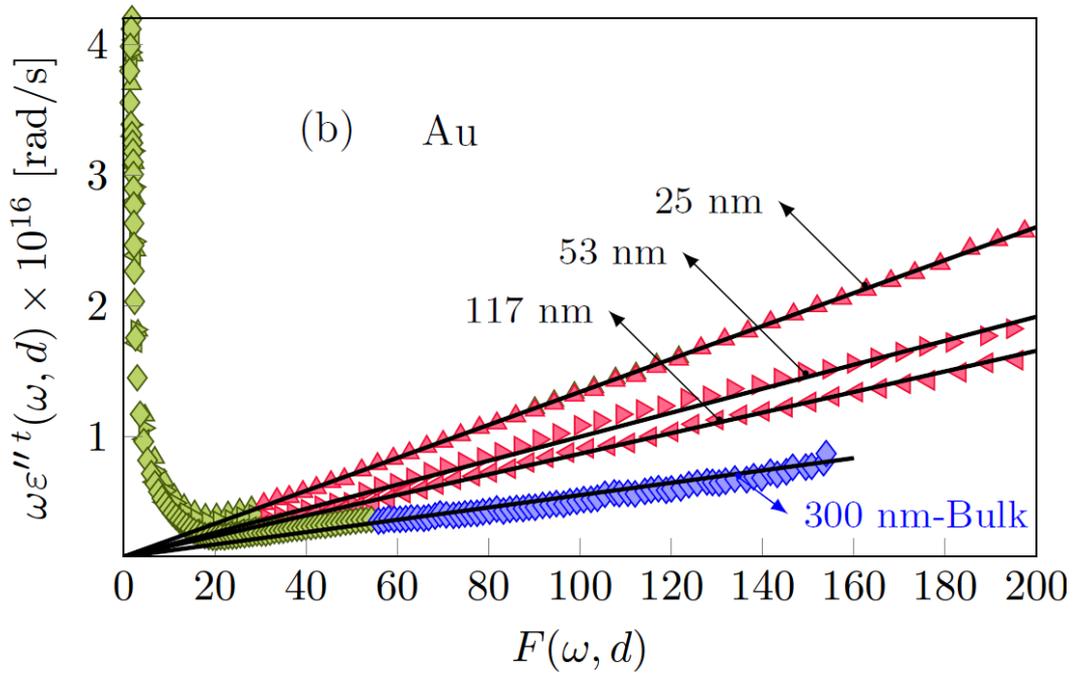

Figure 5: Plot of $\omega\,\varepsilon''^{\,t}(\omega, d)$ vs $F(\omega, d)$ for different Au films thicknesses. Slopes of the linear fits yield to $\gamma_{free}^{t,Au}(d)$ values for each case. Numbers indicate the different thickness values ($d$).

Again, it can be seen that as the film thickness increases, the damping constant decreases. The last line in Table 2 shows a damping constant value calculated from Mc Peak data [31] for 300 nm film thickness, which can be considered as a bulk thickness.

Table 2: Calculated damping constant $\gamma_{free}^{t,Au}(d)$ for different Au film thicknesses.

| $d$ [nm] | $\gamma_{free}^{t,Au}(d) \times 10^{14}$ [rad/s] | $\Delta\gamma_{free}^{t,Au} \times 10^{14}$ [rad/s] |
|---|---|---|
| 4.1 | 2.281 | ±0.217 |
| 6.1 | 1.783 | ±0.166 |
| 9.0 | 1.711 | ±0.216 |
| 25 | 1.322 | ±0.114 |
| 53 | 1.006 | ±0.109 |
| 117 | 0.866 | ±0.102 |
| 300 (Bulk) | 0.532 | ±0.100 |

The experimental data from Table 2 is plotted in Figure 6 in blue symbols. The fit of these data was carried out using Equation (3) with $v_F = 1.38 \times 10^{14} rad/s$ [38]. Again, Fuchs model is used to determine the optimum $l_c$.

The best fit is obtained for $s = 0.11$ and $\gamma_{free}^{Au} = 0.630 \times 10^{14} rad/s$. Again, like in case of Ag, the latter value is very close to that obtained by using Mc Peak data for 300 nm (bulk), as can be observed in the last line in Table 2. With respect to the $s$ parameter, electron scattering at the film boundary is again mainly diffusive (inelastic).

The electron mean free path for bulk Au can be derived similarly to the case of Ag, giving a value of $l_\infty = 22$ nm. This result is consistent with the range of values 22 - 33 nm reported by Bell [40] and Weilmeier et al. [41] using ballistic electron emission microscopy and the value obtained by Nguyen-Truong [42] through GW+T ab initio calculations (29.7 nm).

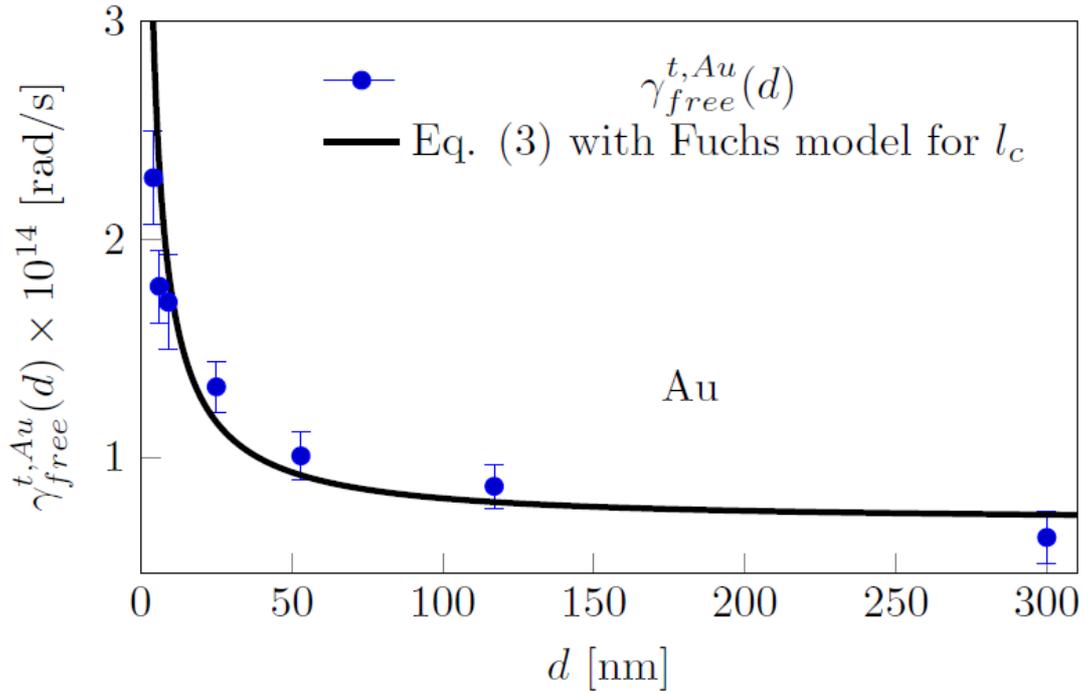

Figure 6. Damping constant in terms of the Au film thickness. Black full line corresponds to a fit using Eq. (3) together with Fuchs model for $l_c$

Table 3, summarizes the values of the bulk damping constant for Ag and Au as derived from the asymptotic values of the fitting curves in Figures 4 and 6, as well as values of $s$ and $l_\infty$.

Table 3. Bulk damping constant, scattering parameter $s$ and bulk mean free path $l_\infty$ for Ag and Au

| Metal | $\gamma_{free} \times 10^{14}$ [rad/s] | $s$ | $l_\infty$ [nm] |
|---|---|---|---|
| Ag | 0.243 | 0.25 | 57 |
| Au | 0.630 | 0.11 | 22 |

## 3.3 Thickness dependent plasma frequency for Ag and Au films

Values of $\omega_p^{t,Ag}(d)$ for Ag films are obtained by using the refractive index values for different thicknesses ($d = 5.0, 6.20, 7.50, 9.80, 12.5$ and $15.0$ nm) given by Zhao et. al. [30]. Notice that, since Equations (7) to (10) are coupled equations, $F(\omega, d)$ range considered (and consequently the wavelength range) for the analysis of plasma frequency must be coincident with that determined for the damping constant $\gamma_{free}^t(d)$, for each thickness and metal analyzed. As an example, Figure 7 shows $G(\omega, d)$ vs $F(\omega, d)$ (Equation (8)) for 5.0 nm Ag thickness. Using the above criterion, the region described by the Drude model is represented by pink dots. From the slope of the linear fit of this region (black full line), plasma frequency can be determined as $\omega_p^{t,Ag}(5.0nm) = 1.286 \times 10^{16} rad/s$.

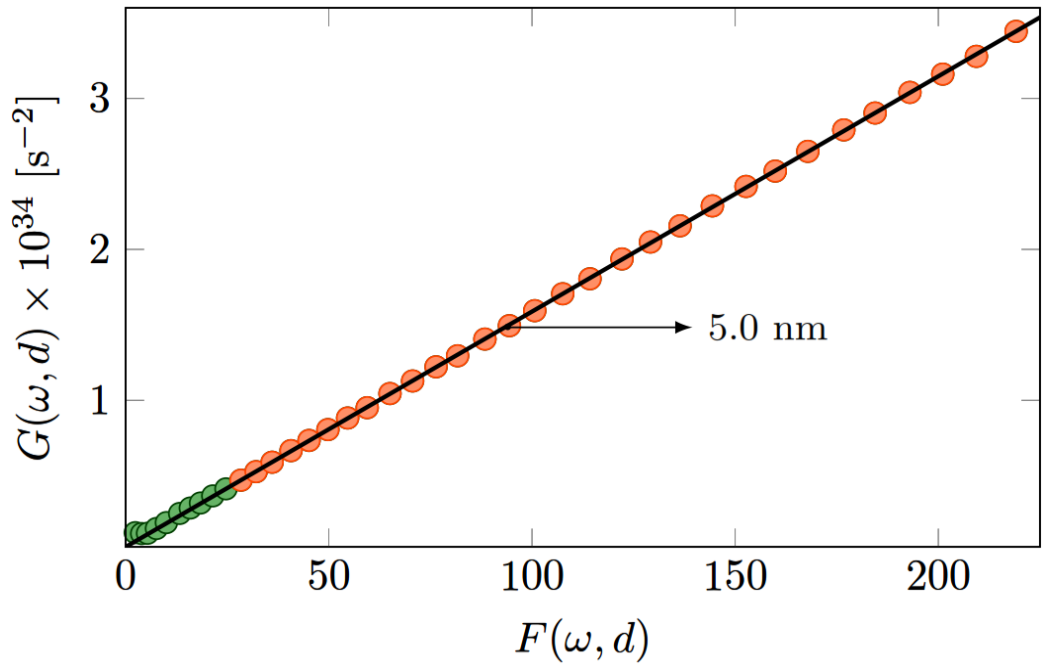

Figure 7: $G(\omega)$ vs $F(\omega)$, for $d = 5.0$ nm thickness Ag film for determining $\omega_p^{t,Ag}$. Black full line is the linear fit of the data shown by pink dots

Figure 8 shows similar analysis for all Ag films thicknesses analyzed (pink dots) together with the experimental bulk 300 nm data by Mc Peak [31] (blue dots). It can be seen that the curves for different thickness have slopes very similar to each other suggesting that plasma frequency is slightly dependent on thickness. The inset is an enlargement of these curves for high values of $F(\omega, d)$, showing this mild dependence. Again, the slope of the linear fit (black full line) can be used to determine $\omega_p^{t,Ag}(d)$ for each thickness, which are summarized in Table 4, together with the errors from the theoretical fits. The last line shows the value of $\omega_p^{t,Ag}$ for bulk thickness.

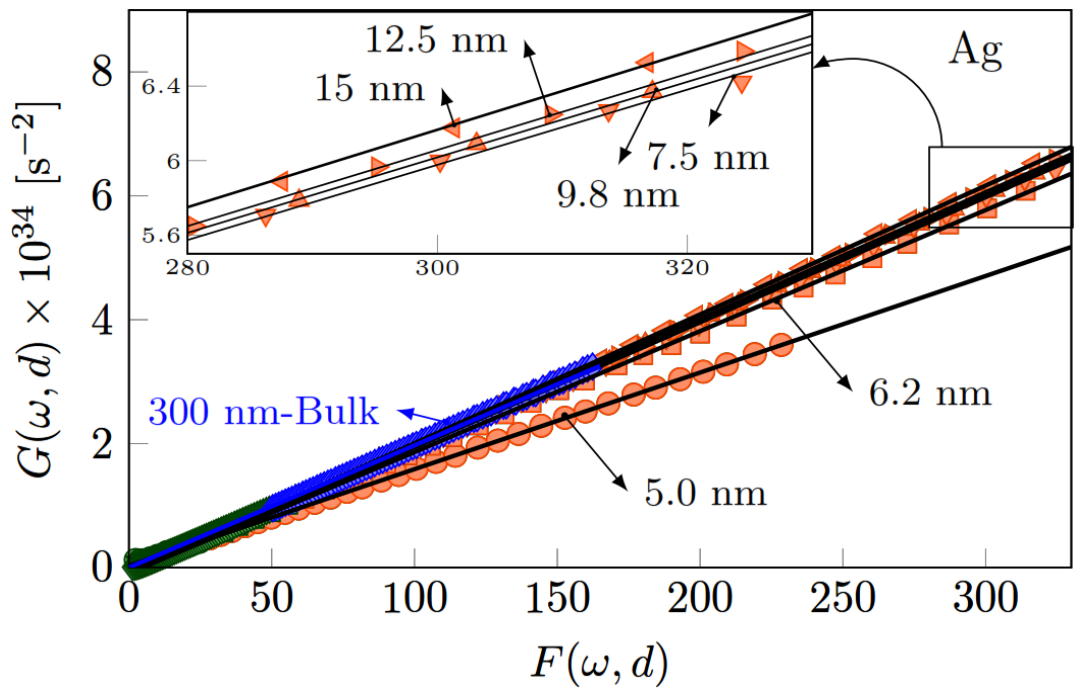

Figure 8: $G(\omega, d)$ vs $F(\omega, d)$ for different Ag films thicknesses. Black straight lines can be used to determine $\omega_p^{t,Ag}(d)$. Inset shows an enlargement of the central region. The numbers indicate the different thickness values ($d$).

Table 4. Calculated $\omega_p^{t,Ag}(d)$ for different film thicknesses.

| $d$[nm] | $\omega_p^{t,Ag} \times 10^{16}$ [rad/s] | $\Delta\omega_p^{t,Ag} \times 10^{15}$ [rad/s] |
|---|---|---|
| 5.0 | 1.286 | ±0.471 |
| 6.2 | 1.381 | ±0.519 |
| 7.5 | 1.407 | ±0.529 |
| 9.8 | 1.413 | ±0.531 |
| 12.5 | 1.417 | ±0.533 |
| 15.0 | 1.430 | ±0.538 |
| 300 (Bulk) | 1.414 | ±0.582 |

It is possible to observe that $\omega_p^{t,Ag}(d)$ shows a redshift as the film thickness decreases for values below 15 nm. This fact is in agreement with the behavior mentioned by Bondarev et al. [25] and Shah et al. [26].

A similar analysis was performed for Au films. Figure 9 shows $G(\omega,d)$ vs $F(\omega,d)$ for different Au films thicknesses, including a 300 nm bulk thickness film (blue dots). Again, $\omega_p^{t,Au}(d)$ can be determined from the slope of the linear fit (black full line). Numbers indicate the corresponding thicknesses. The obtained values of $\omega_p^{t,Au}(d)$ are summarized in Table 5, together with the errors from the theoretical fits. The last line shows the value of $\omega_p^{t,Au}$ for sufficiently large thickness (bulk), obtained by using the refractive index data from Mc Peak [31].

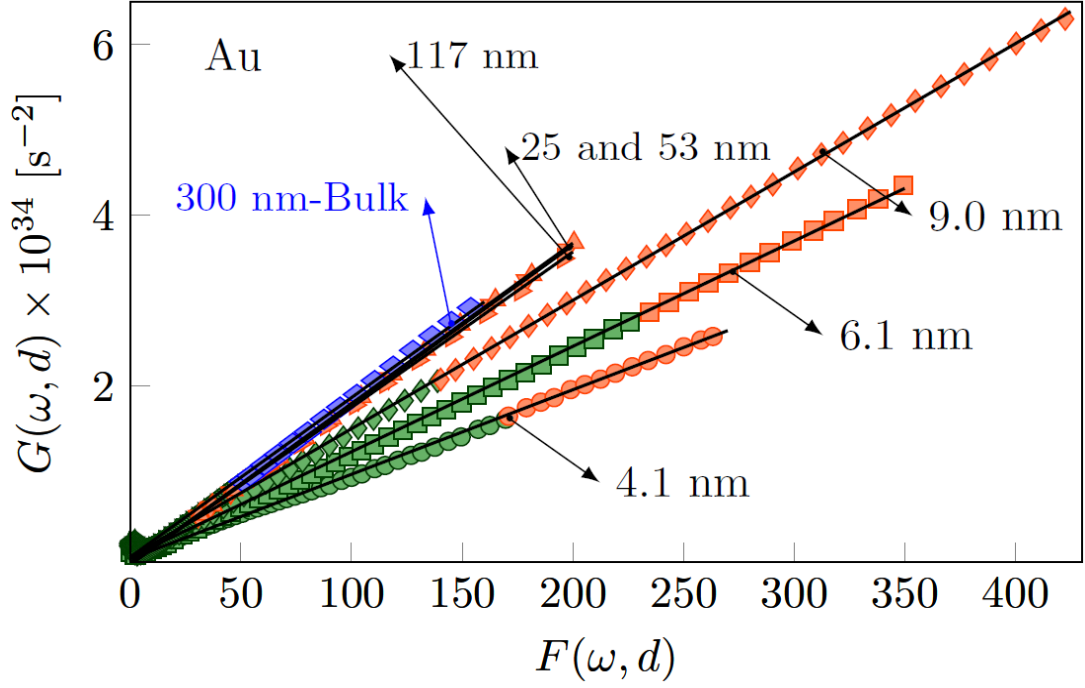

Figure 9: $G(\omega, d)$ vs $F(\omega, d)$ for different Au films thicknesses. Black straight lines can be used to determine $\omega_p^{t,Au}(d)$. Symbols indicate different thickness values ($d$).

Table 5. Calculated $\omega_p^{t,Au}(d)$ for different Au film thicknesses.

| $d$[nm] | $\omega_p^{t,Au} \times 10^{16}$ [rad/s] | $\Delta\omega_p^{t,Au} \times 10^{15}$ [rad/s] |
| --- | --- | --- |
| 4.1 | 0.984 | ±0.284 |
| 6.1 | 1.111 | ±0.201 |
| 9.0 | 1.226 | ±0.231 |
| 25 | 1.341 | ±0.353 |
| 53 | 1.350 | ±0.355 |
| 117 | 1.328 | ±0.350 |
| 300 (Bulk) | 1.358 | ±0.418 |

Here again, $\omega_p^{t,Au}(d)$ shows a redshift as film thickness decreases below 25 nm, in accordance with the general results derived by Bondarev et al [25].

Figure 10 show the experimental values of $\omega_p^{t,Ag}(d)$ (blue circles) and $\omega_p^{t,Au}(d)$ (red squares), together with the corresponding error bars and the fitting curve (dashed and full black lines) of the form:

$$\omega_p^t(d) = \omega_p [1 - e^{-d/d_0}] \qquad (13)$$

where $d_0$ represents a thickness value for which $\omega_p^t(d)$ reaches approximately 63% of the bulk value ($\omega_p$). The best fits corresponds to $d_0$ and $\omega_p$ values shown in Table 6. For Ag, values of $\omega_p^t(d)$ tend asymptotically to the bulk value for thicknesses larger than 10 nm, while for Au, this behavior is reached for thicknesses larger than 25 nm.

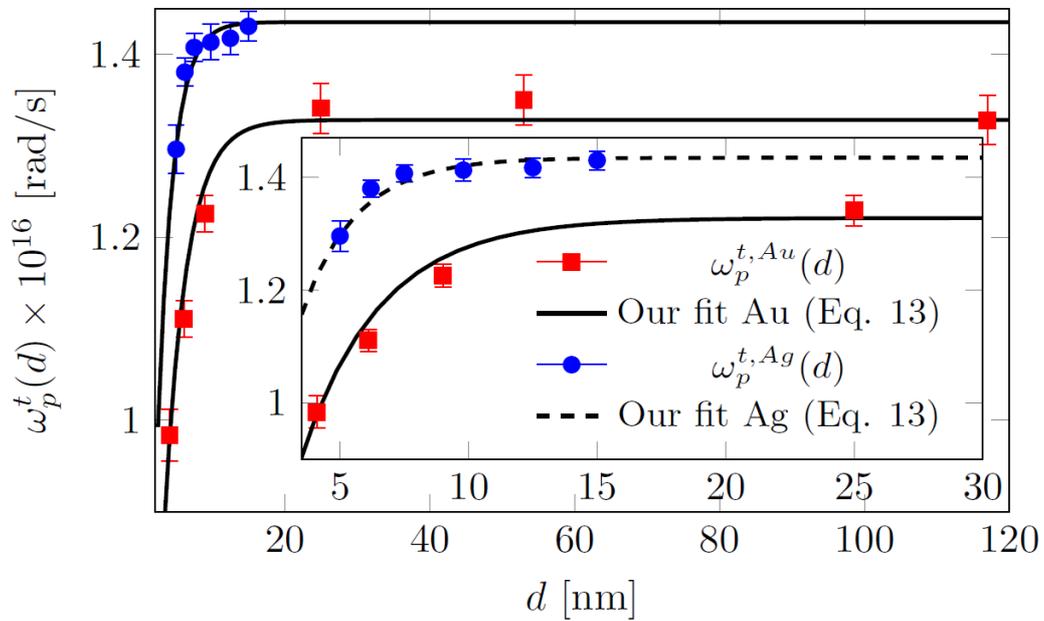

Figure 10. $\omega_p^{t,Ag}(d)$ and $\omega_p^{t,Au}(d)$ in terms of film thickness. Blue circle and red square symbols are experimental values corresponding to Ag and Au, respectively. Dashed and full black lines correspond to the fitting curve for Ag and Au, respectively, obtained by using Eq. (13).

Table 6: Values for $d_0$ and $\omega_p$ for Ag and Au

| Metal | $d_0$ [nm] | $\omega_p \times 10^{16}$ [rad/s] |
|-------|------------|-----------------------------------|
| Ag    | 2.125      | 1.435                             |
| Au    | 3.065      | 1.328                             |

So, Ag and Au $\omega_p$ values included in Table 6 are statistically meaningful bulk values of plasma frequency and will be used in section 3.4 to rewrite the thickness dependent dielectric function.

**3.4 Thickness dependent dielectric function expression for Ag and Au films**

Having found the values for $\gamma^t_{free}(d)$ and $\omega^t_p(d)$ for Ag and Au (Tables 1, 2, 4 and 5), we can now introduce them in Eq. (5) to reproduce the complex $\varepsilon^t(\omega, d)$ within the Drude model approximation. Figure 11 shows the real ($\varepsilon'^{t, Ag}(\omega, d)$, red, right axis) and imaginary ($\varepsilon''^{t, Ag}(\omega, d)$, blue, left axis) parts of the wavelength dependent dielectric function for Ag films. The experimental values for different film thicknesses are indicated with symbols. Bulk data is plotted in gray circles and its fitting curve in black full line.

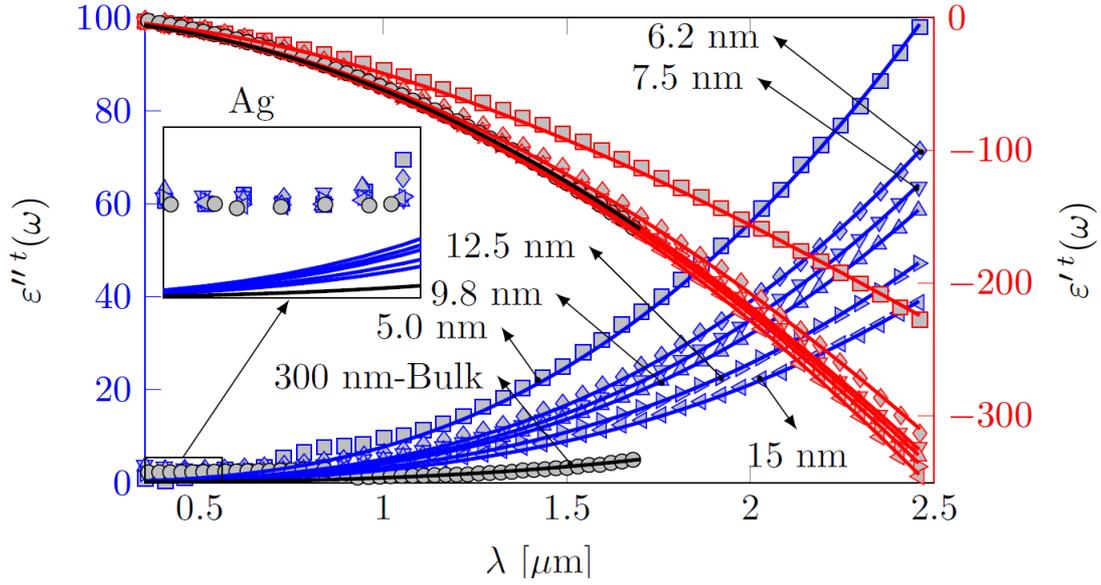

Figure 11. Real (red to the right) and imaginary (blue to the left) part of Ag dielectric function vs wavelength for different thicknesses ($d$). The fitting curves corresponding to different thicknesses are indicated by numbers and symbols.

The fitting of the experimental data for different thicknesses using Equation (5) are represented by full lines (red for the real part and blue for imaginary part). It can be seen that there is a very good fitting of the experimental data for the wavelength range larger than about 1.25 μ$m$, but a noticeable deviation for the short wavelength tail of the curve (inset), as expected since we were working only within the Drude model and bound electron contribution is not considered in this approach.

A similar analysis was performed for Au films. Figure 12 (a) and (b) shows the real ($\varepsilon'^{t,Au}(\omega,d)$, red, right axis) and imaginary ($\varepsilon''^{t,Au}(\omega,d)$, blue, left axis) parts of the wavelength dependent dielectric function for Au films. Bulk data is plotted in gray circles and its fitting curve in black full line. Panel (a) shows the results corresponding to 25 nm, 53 nm, 117 nm and bulk (300 nm) thicknesses. Panel (b) shows the results corresponding to 4.1 nm, 6.1 nm, 9 nm and bulk (300 nm) thicknesses. It is noticeable in panel (a) a clear deviation from Drude model for the short wavelength of the experimental data, similarly to the behavior for Ag, due to the lack of consideration of the bound electron contribution within Drude model. Another fact to be observed is that, for both metals, bound electron contribution for the imaginary part of the dielectric function is more noticeable than that for the real part. Inspection of the short wavelength range (λ < 600 nm) in Figures 11 and 12 shows that this deviation is almost the same for the different film thicknesses analized, including the 300 nm thickness data, considered bulk. This fact suggests that bound electron contribution may be safely considered independent of thickness and may be taken equal to the bulk film contribution for the range of studied thicknesses. This behavior is in agreement with the findings by Scaffardi and Tocho [13] for the case of tiny spherical nanoparticles.

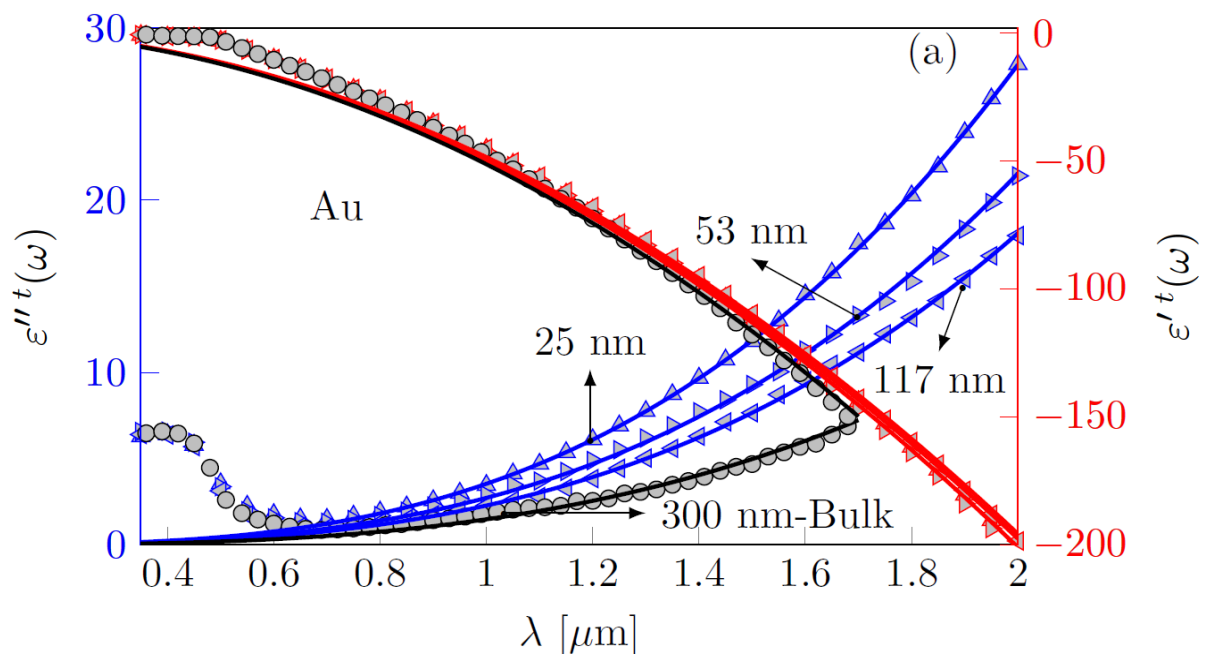

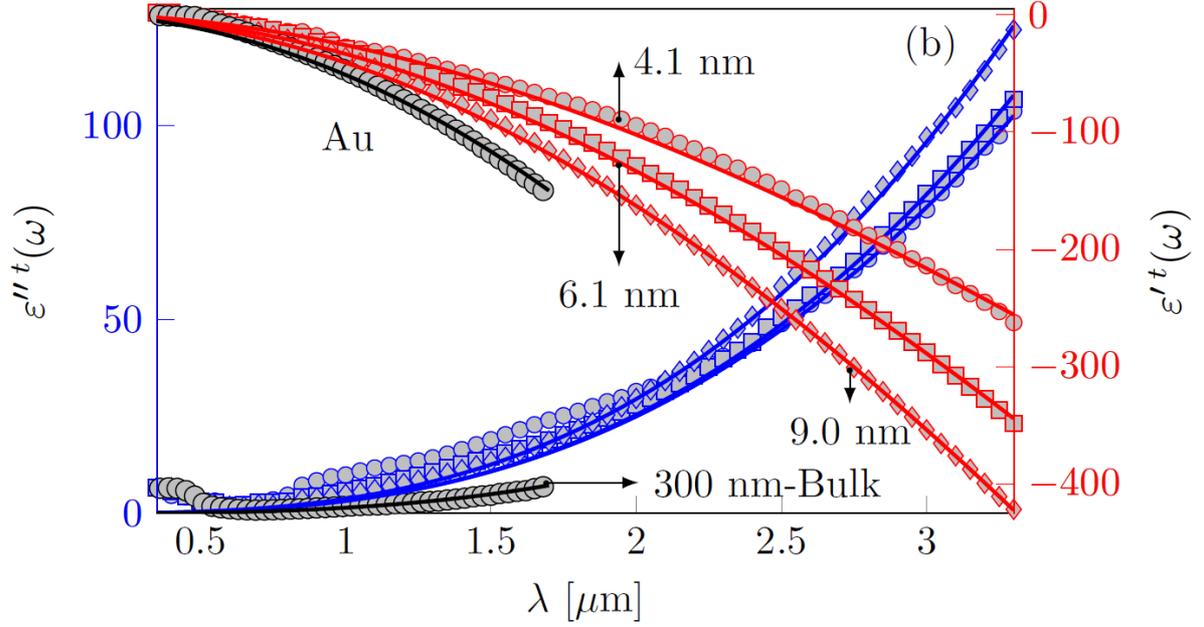

Figure 12. Real (red to the right) and imaginary (blue to the left) part of Au dielectric function vs wavelength for different thicknesses (d). The curves corresponding to different thicknesses are indicated by numbers and symbols.

With this consideration, the full final expression of the thickness dependent dielectric function, following Equation (1), can be rewritten in the form:

$$\varepsilon^t(\omega, d) = \varepsilon_{bulk}^{free,t}(\omega, d) + \varepsilon_{bulk}^{bound}(\omega) \tag{14}$$

and replacing Equation (5) into (14):

$$\varepsilon^t(\omega, d) = 1 - \frac{\left(\omega_p^t(d)\right)^2}{\omega^2 + i\omega\gamma_{free} + i\omega\frac{v_F}{l_c}} + \varepsilon_{bulk}^{bound}(\omega) \tag{15}$$

In general in the literature $\varepsilon_{bulk}^{bound}$ is considered independent of frequency, and taken into account by adding a constant value $\varepsilon_\infty$ in the expression of the Drude model, which does not represent an adequate approximation. Our approach is to consider the frequency dependent bound electron contribution ($\varepsilon_{bulk}^{bound}(\omega)$) directly from experimental measurement of the bulk dielectric function. Then, introducing the expression $\varepsilon_{bulk}^{bound}(\omega) = \varepsilon_{bulk}(\omega) - \varepsilon_{bulk}^{free}(\omega)$ from Equation (1), where $\varepsilon_{bulk}(\omega)$ is the frequency dependent bulk experimental data and $\varepsilon_{bulk}^{free}(\omega)$ is taken from Equation (2), Equation (15) can be rewritten as:

$$\varepsilon^t(\omega, d) = \varepsilon_{bulk}(\omega) + \frac{\left(\omega_p\right)^2}{\omega^2 + i\omega\gamma_{free}} - \frac{\left(\omega_p^t(d)\right)^2}{\omega^2 + i\omega\gamma_{free} + i\omega\frac{v_F}{l_c}} \qquad (16)$$

So, thickness-dependent dielectric function can be described, based on Equation (16), as a function of the experimental bulk dielectric function $\varepsilon_{bulk}(\omega)$, and the previously determined $\gamma_{free}$ (Tables 3), $\omega_p$ (Tables 6), $l_c$ (Equation 11) and $\omega_p^t(d) = \omega_p(1 - e^{-d/d_0})$ (Equation 13):

$$\varepsilon^t(\omega, d) = \varepsilon_{bulk}(\omega) + \frac{\left(\omega_p\right)^2}{\omega^2 + i\omega\gamma_{free}} - \frac{\left(\omega_p(1-e^{-d/d_0})\right)^2}{\omega^2 + i\omega\gamma_{free} + i\omega\frac{v_F}{l_c}} \qquad (17)$$

Equation [17] contains all the necessary information to allow calculating the full and exact complex dielectric function for Ag and Au thin films with arbitrary thicknesses $d$. Notice that this calculation can be carried out taking the experimental bulk dielectric function and adding thickness dependent correction terms.

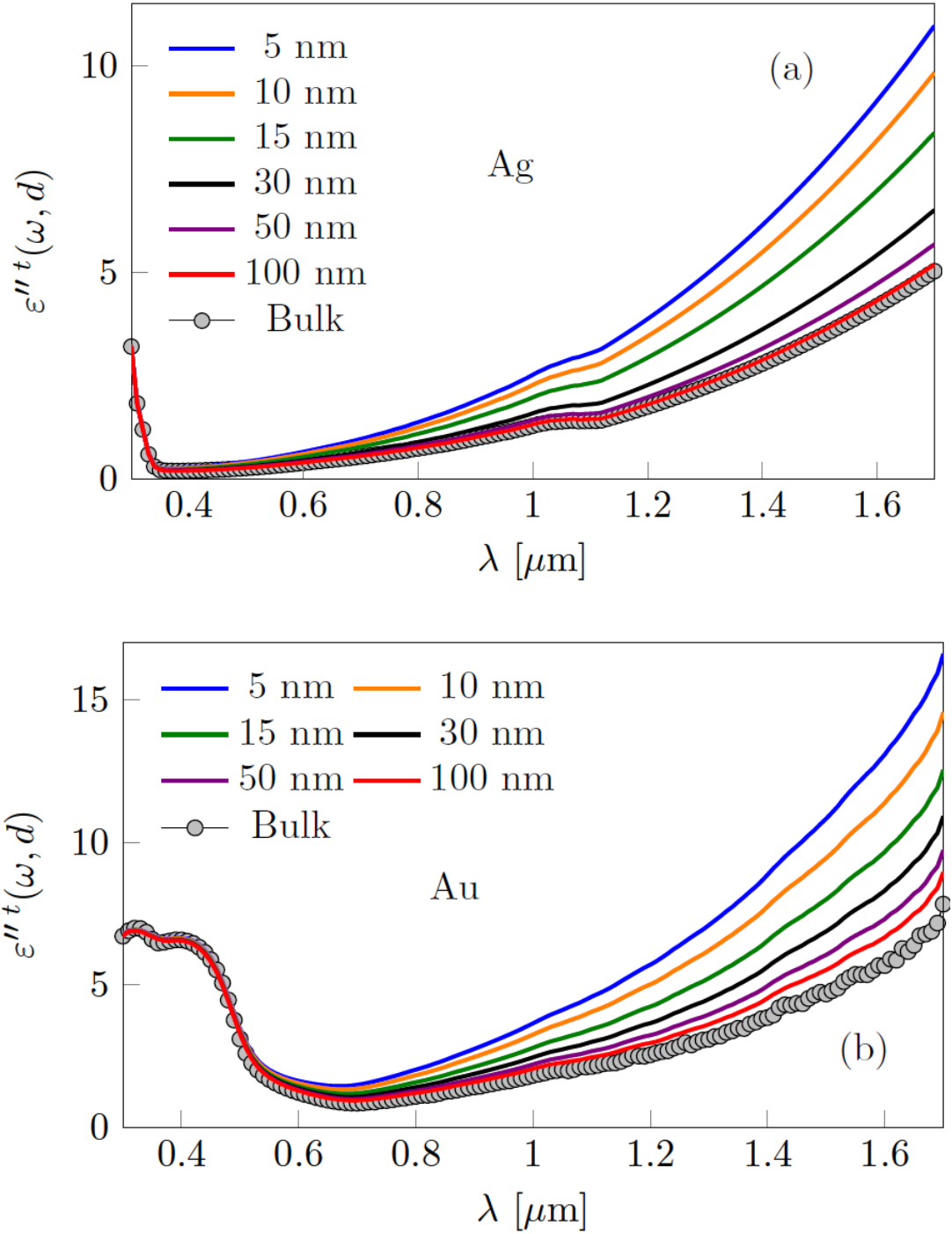

Figure 13. Imaginary part $\varepsilon'''^{t}(\omega, d)$ vs wavelengths for different thicknesses. (a) Ag; (b) Au. Bulk curves (gray circles) correspond to Mc Peak data [30].

To show an example of this calculation and recalling that the imaginary part of the dielectric function has a larger bound electron contribution than the real part, Figures 13 (a) and (b) show respectively, the imaginary part of the dielectric function given by Equation (17) for Ag and Au vs wavelength for a given film thicknesses $d$. Experimental bulk data are included in gray circles and the calculated dielectric function for different thin films, in different color lines. It can be observed that as the thickness increases, the curves tend to the bulk value.

**Conclusions**

From ultrathin (few atomic monolayers) Ag and Au films experimental refractive index data and using a modified Drude model, we determined without approximation for the first time, the thickness dependence of the plasma frequency and damping constant for Ag and Au plasmonic metals films. Our general results agree with the behavior theoretically predicted by Bondarev [25].

The dependence of the damping constant with film thickness was fitted using Fuchs model for $l_c$, taking $\gamma_{free}$ and $s$ as the fitting parameters. The best fits for Ag and Au were obtained for an $s$ value that corresponds to electron scattering at the film boundary being mainly diffusive (inelastic). For Ag, $\gamma_{free}^t(d)$ tends asymptotically to its bulk value ($\gamma_{free}^{Ag} = 0.243 \times 10^{14} rad/s$) for thicknesses larger than 50 nm, while for Au this limit ($\gamma_{free}^{Au} = 0.630 \times 10^{14} rad/s$) is reached for thicknesses larger than 100 nm. Our method allow us to determine the mean free path $l_\infty$, obtaining 57 nm (for Ag) and 22 nm (for Au).

The asymptotic dependence of the plasma frequency with increasing film thickness was satisfactorily fitted using Equation (13) for all the experimental data. For Ag, values of $\omega_p^t(d)$ tend to its bulk value ($\omega_p^{Ag} = 1.435 \times 10^{16} rad/s$) for thicknesses larger than 10 nm, while for Au this limit ($\omega_p^{Au} = 1.328 \times 10^{16} rad/s$) is reached for thicknesses larger than 15 nm.

For Ag as well as for Au, plasma frequency tends to bulk values for smaller thicknesses compared with the damping constant.

According to the results obtained in this work, it is important to notice that care should be taken when selecting refractive index values that correspond to bulk situations. In this sense the values in the literature used for determining the optical constants should correspond to films with thickness over 100 nm. For example, the refractive index values for a Ag and Au reported by McPeack [31] for thickness of 300 nm may be well considered as bulk.

Finally, based on the expressions derived in this work for the film thickness dependence of the damping constant and plasma frequency, we calculate for the first time the complex dielectric function for a thin film as the sum of the bulk dielectric function plus two additive terms containing the film thickness dependence.

**Acknowledgments**

This work was granted by PIP CONICET 0279, MINCyT PME 2006-00018, 11/I237, 11/I239 of Facultad de Ingeniería at Universidad Nacional de La Plata (UNLP), PICT2018-04558 and PICT A 00074 of Agencia Nacional de Promoción Científica y Tecnológica, MINCyT, Argentina.

**Conflict of Interest**

The authors declare no conflict of interest.